\documentclass[pre,twocolumn,amsmath,amssymb,showkeys,nofootinbib,floatfix,superscriptaddress]{revtex4}

\usepackage{graphicx}
\usepackage{dcolumn}
\usepackage{bm}
\usepackage{times}

\begin{document}


\title{Finite-size effects in the 
dependency networks of free and open-source software}

\author{Rajiv Nair}
\email{rajiv@tiss.edu}
\affiliation{Tata Institute of Social Sciences, V. N. Purav Marg,
Deonar, Mumbai 400088, India} 
\author{G. Nagarjuna}
\email{nagarjun@gnowledge.org}
\affiliation{Homi Bhabha Centre for Science Education, 
Tata Institute of Fundamental Research, V. N. Purav Marg, 
Mankhurd, Mumbai 400088, India}
\author{Arnab K. Ray}
\email{arnab.kumar@juet.ac.in}
\affiliation{Department of Physics, Jaypee University of 
Engineering and Technology, Raghogarh, Guna 473226, 
Madhya Pradesh, India}


\date{\today}

\begin{abstract}
We propose a continuum model for the degree distribution 
of directed networks in free and open-source software. The degree 
distributions of links in both the in-directed and out-directed
dependency networks follow Zipf's law 
for the intermediate nodes, but the heavily linked nodes  
and the poorly linked nodes deviate from this
trend and exhibit finite-size effects.
The finite-size parameters make a quantitative 
distinction between the in-directed and out-directed networks.
For the out-degree distribution, the initial 
condition for a dynamic evolution corresponds to 
the limiting count of the most heavily liked nodes that 
the out-directed network can finally have. 
The number of nodes contributing out-directed
links grows with every generation of software release, 
but this growth ultimately saturates towards a terminal value due 
to the finiteness of semantic possibilities in the network.  
\end{abstract}

\maketitle

\section{Introduction}
\label{sec1}

Scale-free distributions in complex networks~\cite{domen2,nbw,
bbv,domen,stronat,albar,new,amot, markn} span across diverse 
domains like the 
World Wide Web~\cite{domen2,ajb} and the Internet~\cite{domen2}, 
social, ecological, biological and linguistic networks~\cite{albar}, 
trade and 
business networks~\cite{acbkc1}, and syntactic and semantic 
networks~\cite{cansolko,cancho,stete}.  
Scale-free features have also been discovered in 
electronic circuits~\cite{canjaso} and in the architecture of 
computer software~\cite{valcaso}. 
The structure of object-oriented software is a heterogeneous
network, characterized by a power-law distribution~\cite{valsol},
and it is on the basis of scale-free networks that software 
fragility is explained~\cite{challom}. 
Power-law features exist in the 
inter-package dependency networks in free and open-source 
software ({\it FOSS})~\cite{labwal}, and studies have 
shown that modifications in this type of a software network also
follow a power-law decay in time~\cite{chaldu,chalval}. 

Continuing on the software theme, while installing
a software package from the {\it Debian GNU/Linux} distribution, 
many other packages, known as the ``dependencies", are needed
as prerequisites. This leads to a dependency-based 
network among all the packages, with each of these packages
being a node in a network of 
dependency relationships. Each dependency relationship connecting 
any two packages (nodes) is a link (an edge), and every
link establishes a relation between a prior package and a posterior 
package, whereby the functions defined in the prior package are 
invoked in the posterior package. So what emerges is 
a semantic network, with a directed flow of meaning, 
determined by the direction of the links. 

Semantic networks are a subject of major interest, especially 
where small-world structures~\cite{watstro} and scale-free 
aspects~\cite{albar} of networks are concerned~\cite{stete}. 
With particular regard to component-based software, a semantic 
relationship among components underlies
the network of what are known as strong dependencies~\cite{zack}. 
The components of a {\it FOSS} network are 
interconnected by various relationships (including a negative
one, ``Conflicts"), and only one of these is based on the 
field, ``Depends". This again is further categorized into the 
two cases of strong dependencies and direct dependencies, with 
a correlation between the two cases~\cite{zack}. As regards
direct dependencies, the scale-free character of the 
{\it Debian GNU/Linux} distribution has been 
studied~\cite{labwal,mssk}. 

In our study, the semantic network of nodes in the {\it Debian} 
distribution is founded on one single principle 
running through all the nodes: $Y$ depends 
on $X$; its inverse, $X$ is required for $Y$. 
The semantic network so formed is a straightforward 
dependency-based directed network only. 
Considering any particular node in such a directed network, 
its links (the relations with other 
nodes) are of two types, incoming links and 
outgoing links, as a result of which, there will arise two 
distinct types of directed network~\cite{albar}. For the network 
of incoming links in the {\it Etch} release of {\it Debian},
one study~\cite{mssk} has empirically tested Zipf's law in 
the {\it GNU/Linux} distribution, a phenomenon that,
discovered originally in the occurrence frequency of words 
in natural languages~\cite{zipf}, has over the years emerged
widely in many other areas.   

Our work affirms the existence
of Zipf's law as a universal feature underlying the {\it FOSS} 
network. Here, in fact, both the networks of incoming and outgoing 
links follow Zipf's law. However, simple 
power-law properties do not suffice to provide a complete global 
model for directed networks. For
any system with a finite size, the power-law trend is not manifested 
indefinitely~\cite{barstan,nsw}, and for a {\it FOSS} 
network, this matter awaits a 
thorough investigation~\cite{mssk}. Deviations from the power-law trend 
appear for both the profusely-linked and the sparsely-linked nodes. 
The former case corresponds to the distribution of a disproportionately 
high number of links connected to a very few important nodes (the
so-called ``hubs" or rich nodes or top nodes).  
The particular properties of all these 
outlying nodes, as well as any distinguishing characteristic 
of the two directed networks, can only be known by studying  
the finite-size effects (equivalently the saturation properties) 
in the respective networks~\cite{nnrjphys},
and by understanding how these effects are related to the 
semantic structure in the network.  
These are the principal objects of our investigation. 

\section{A nonlinear continuum model}
\label{sec2}

The main advantage of the {\it Debian GNU/Linux} distribution 
is that it is the largest component-based system that can be
accessed freely for study~\cite{zack}. The mathematical modelling 
of this {\it FOSS} network has been carried 
out here primarily with the help of data collected from the two 
stable {\it Debian} releases, {\it Etch} ({\it Debian GNU/Linux 4.0}
and {\it Lenny} 
({\it Debian GNU/Linux 5.0}).\footnote{\tt {http://www.debian.org/releases}}
The networks of both the incoming links and the outgoing links span 
about $18000$ packages (nodes) in the {\it Etch} release, 
while in the {\it Lenny} release, the corresponding number of 
packages is about $23000$. For this work, the 
chosen computer architecture supported by both the releases
is {\it AMD64}. The dynamic features of the model have further 
been grounded on the first three
generations of {\it Debian releases}, i.e. 
{\it Buzz} ({\it Debian GNU/Linux 1.1}), 
{\it Hamm} ({\it Debian GNU/Linux 2.0}) and
{\it Woody} ({\it Debian GNU/Linux 3.0}), all of which are 
supported by the architecture {\it i386}. 
The model founded with the help of the {\it Etch} and {\it Lenny}
releases, shows a retrospective compatibility 
with the earlier releases, and moving forward in time, it is also 
in consonance with the features shown by the latest stable
{\it Debian} release, {\it Squeeze} ({\it Debian GNU/Linux 6.0}), 
which is again based on the {\it AMD64} architecture.   
The graphical results presented in this paper 
are based mostly on the 
three latest releases, {\it Etch}, {\it Lenny} and {\it Squeeze}.
All of these releases have a substantial number of nodes and links, 
and even though these numbers are to be counted only discretely, 
the largeness of their total count allows a continuum description 
to be adopted, using a differential equation.

For developing the model we need to count
the actual number of software packages, $\phi$, which are connected 
by a particular number of links, $x$, in either kind of network.  
This gives an unnormalized frequency distribution 
of $\phi \equiv \phi (x)$ versus $x$. Normalizing this 
distribution in terms of the relative frequency distribution of 
the occurrence of packages would have yielded the usual probability
density function. To provide a continuum model for any power-law
feature in this frequency distribution, we posit a 
nonlinear logistic-type equation,
\begin{equation}
\label{logis}
\left(x + \lambda\right) \frac{{\mathrm d}\phi}{{\mathrm d}x} =
\alpha \phi \left(1 - \eta \phi^{\mu} \right) \,,
\end{equation}
in which $\alpha$ is a power-law exponent, $\mu$ is a nonlinear
saturation exponent, $\eta$ is a ``tuning"
parameter for nonlinearity and $\lambda$ is another parameter 
that is instrumental in setting a limiting scale for the 
poorly connected nodes. The motivation behind this mathematical
prescription can be easily followed by noting that when 
$\eta = \lambda = 0$, there will be a globally 
valid power-law distribution. 
However, when the distribution is finite, the power-law trend fails 
to hold true beyond intermediate scales of $x$. Such deviations 
from a full power-law behaviour is especially prominent for high 
values of $x$ (related to the very heavily connected nodes) and, 
therefore, it can be argued that finiteness in the 
distribution is closely related to its saturation. This type 
of saturation behaviour is frequently modelled by a nonlinear 
logistic equation~\cite{montroll,stro}, and 
so, to understand the saturation properties of the highly 
connected nodes in the {\it Debian} network, it will be necessary
to understand the part played by nonlinearity.

Integration of Eq.~(\ref{logis}), which is a nonlinear differential 
equation, is done by making suitable substitutions on  
$\phi^\mu$ and $x + \lambda$, followed by the application 
of partial fractions. After that we get the 
integral solution of Eq.~(\ref{logis}) as (for $\mu \neq 0$) 
\begin{equation}
\label{integ}
\phi (x) = \left[\eta +
\left(\frac{x+\lambda}{c}\right)^{-\mu \alpha}\right]^{-1/\mu} \,, 
\end{equation}
where $c$ is an integration constant. Evidently, 
when $\eta = \lambda = 0$ (with the former condition implying the 
absence of nonlinearity), there will be a global power-law 
distribution, going as $\phi (x) = (x/c)^{\alpha}$, 
regardless of any non-zero value of $\mu$. The situation becomes 
quite different, however, when both $\eta$ and $\lambda$ have non-zero
values. In this situation, the network will exhibit a saturation 
behaviour on extreme scales of $x$ (both low and high). For high
values of $x$, this can be easily appreciated from Eq.~(\ref{logis}) 
itself, wherefrom the limiting value of $\phi$ is obtained as 
$\phi = \eta^{-1/\mu}$. 

\section{Model fitting of the {\it FOSS} network}
\label{sec3}

\begin{figure}[floatfix]
\begin{center}
\includegraphics[scale=0.65, angle=0]{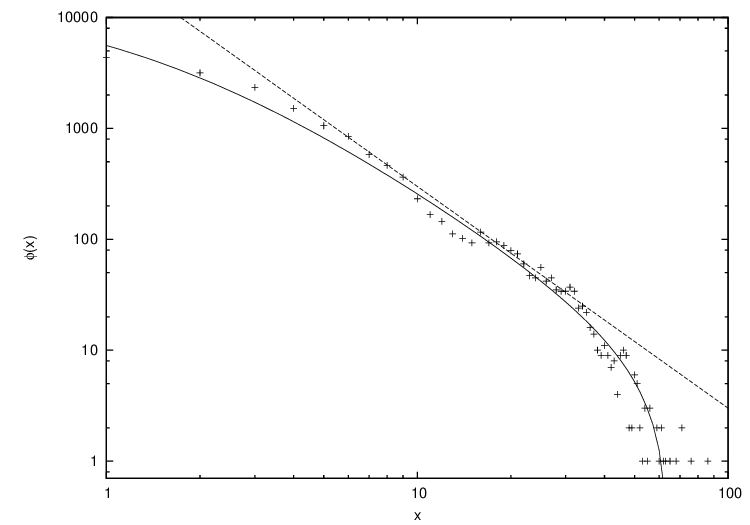}
\caption{\label{f1}\small{For the network of incoming links in the
{\it Etch} release, the degree distribution
shows a good fit in the intermediate region with a
power-law exponent, $\alpha =-2$ (as indicated by the dotted straight
line), which validates Zipf's law. However, for large values of $x$,
there is a saturation behaviour towards a limiting scale that is
modelled well with the parameter, $\eta = -8$. 
When $x$ is small, the fit is good for $\lambda = 1.5$. The global
fit becomes possible only when $\mu = -1$, which turns out to be a
universally valid number. For this specific plot, the data are
fitted by $c \simeq 190$.}}
\end{center}
\end{figure}
\begin{figure}[floatfix]
\begin{center}
\includegraphics[scale=0.65, angle=0]{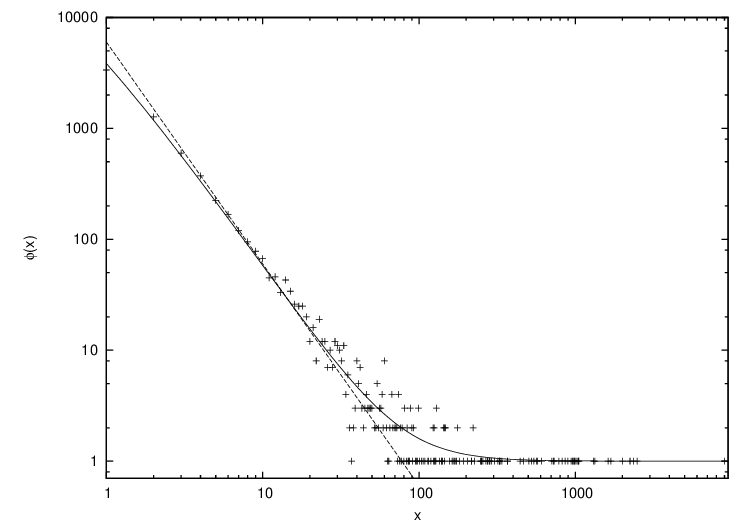}
\caption{\label{f2}\small{For the network of outgoing links in 
the {\it Etch} release, the
degree distribution of intermediate nodes is again
modelled well by a power-law exponent, $\alpha = -2$, which is Zipf's
law (as the dotted straight line shows). However, the saturation
behaviour of the top nodes is different from that of the network
of incoming links. There is a clear convergence of $\phi$ towards a
limit given by $\eta = 1$ (with $\mu$ remaining unchanged at $-1$).
For the poorly linked nodes the convergence is attained for
$\lambda = 0.25$. Thus, when $\alpha$ and $\mu$ remain the same,
the value and the sign of $\eta$, as well as the value of $\lambda$,
distinguish the type of a dependency network.
The data are fitted for $c \simeq 80$.
A solitary top node is to be seen for $x = 9025$.}}
\end{center}
\end{figure}
\begin{figure}[floatfix]
\begin{center}
\includegraphics[scale=0.65, angle=0]{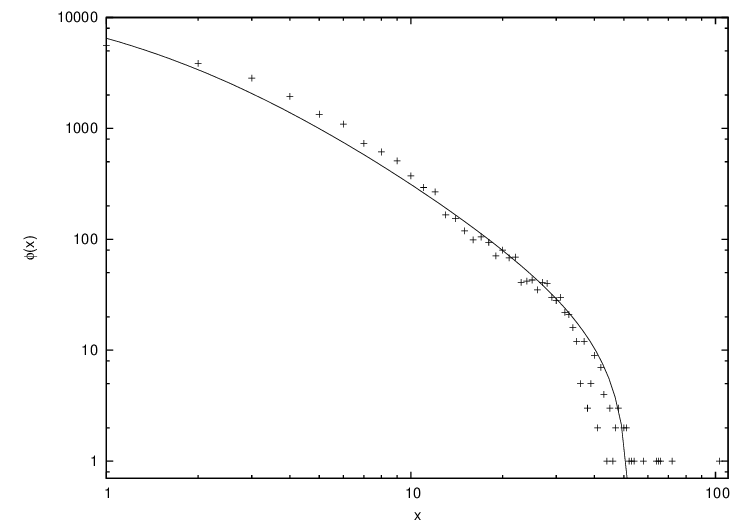}
\caption{\label{f3}\small{For the network of incoming links in the 
{\it Lenny} release, the intermediate nodes (fitted with a power-law
exponent, $\alpha =-2$) uphold Zipf's law once again. For large values
of $x$, however, the saturation behaviour towards a limiting scale
of $\phi$ is modelled by the value, $\eta = -15$. 
When $x$ is small, the fit is good for $\lambda = 1.6$.
Once again $\mu = -1$, but for this particular plot, $c \simeq 210$.
The richly linked nodes here are less
connected than what they are in the case of the {\it Etch} release.}}
\end{center}
\end{figure}
\begin{figure}[floatfix]
\begin{center}
\includegraphics[scale=0.65, angle=0]{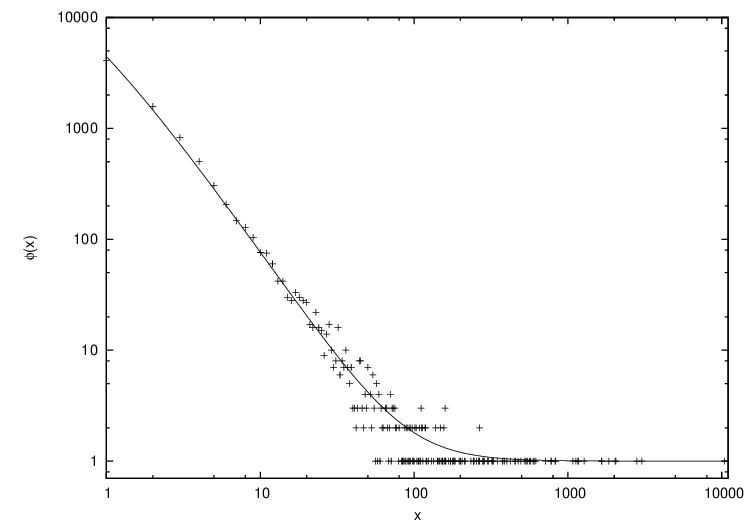}
\caption{\label{f4}\small{For the network of outgoing links in
the {\it Lenny} release, the distribution of intermediate nodes obeys
Zipf's law, as the power-law exponent, $\alpha = -2$, shows. The
saturation behaviour of the top nodes remains the same as it is
for the {\it Etch} data. The convergence of $\phi$ towards a limit
set by $\eta = 1$ is evident, with $\mu = -1$, as before. For
the poorly linked nodes, the convergence is given by $\lambda = 0.35$.
The other value that distinguishes the out-degree distribution in
the {\it Lenny} release, from that in the {\it Etch} release, is
$c \simeq 90$. A solitary top node is to be seen for $x = 10446$.}}
\end{center}
\end{figure}
The parameters $\alpha$, $\mu$, $\eta$, $\lambda$ and $c$ in the 
solution given by Eq.~(\ref{integ}) can now be fixed by the 
distribution of links and nodes in the {\it Debian}
repository. In Fig.~\ref{f1} the degree distribution of 
incoming links in the {\it Etch} release is 
plotted. The dotted straight line in this  
$\log$-$\log$ plot indicates a purely power-law behaviour. While
this gives a satisfactory description of the distribution 
on intermediate scales of $x$, there is a clear departure from the
power law both as $x \longrightarrow 0$ and $x \longrightarrow \infty$. 
The solution given by Eq.~(\ref{integ}) fits the power law, as well 
as the departure from it, at both the small-connectivity and 
the high-connectivity ends. Among all the parameters,  
the values of $\alpha$ and $\mu$ remain unchanged while 
modelling the degree distribution of outgoing links, 
as shown in the plot in Fig.~\ref{f2}. The obvious implication 
of $\alpha = -2$ in both the cases is that Zipf's law universally 
underlies the frequency distribution of the intermediate nodes and 
links in both kinds of network. The only quantitative measures to 
distinguish between the two networks are the values of $\eta$,  
$\lambda$ and $c$. 

Similarly, data from the {\it Lenny} release have been plotted in 
Figs.~\ref{f3} \&~\ref{f4}. The former plot gives the in-degree 
distribution of the nodes, while the latter gives the out-degree 
distribution. The values of $\eta$, $\lambda$ and $c$ in the
in-degree distribution of the {\it Lenny} release change with 
respect to the previous release, {\it Etch}. With changing values 
of these particular parameters, the saturation properties in the
in-degree distribution, therefore, undergo a significant quantitative 
change at the highly connected end. In contrast, for the out-degree 
distribution the changes across a new generation of {\it Debian} 
release are fitted by varying the values of $\lambda$ and $c$.
The fact that $\eta$ remains the same as before while $\lambda$
changes, implies that the saturation properties remain unchanged 
at the richly linked end, but changes at the poorly connected end. 
Changes in the value of $c$ for a particular degree distribution, 
cause a translation of the model curve in the $x$--$\phi$ plane. 
And, as Figs.~\ref{f1},~\ref{f2},~\ref{f3} \&~\ref{f4}
indicate, Zipf's law prevails in all the cases with $\alpha =-2$. 

To appreciate the mathematical implications of 
obtaining $\mu =-1$ from the
data, a power-series expansion of Eq.~(\ref{integ}) has to be
carried out, leading to the infinite series, 
\begin{eqnarray}
\label{series1}
\phi (x) &=& \left(\frac{x+\lambda}{c}\right)^{\alpha}-\frac{\eta}{\mu}
\left(\frac{x+\lambda}{c}\right)^{{\alpha}(\mu + 1)} \nonumber \\
& & + \frac{\mu + 1}{2} \left(\frac{\eta}{\mu}\right)^2
\left(\frac{x+\lambda}{c}\right)^{{\alpha}(2 \mu + 1)}+\cdots \,,
\end{eqnarray}
from which it is not difficult to see that a self-contained and 
natural truncation for this series can only be achieved when 
$\mu = -1$. This is necessary if the scale-free character of the
distribution is to be preserved, otherwise, with $\mu \neq 1$, 
different terms in Eq.~(\ref{series1}) will become dominant on 
different scales of $x$. It is remarkable that the {\it Debian} 
data conform to this fact, and consequently, with $\mu =-1$, 
Eq.~(\ref{logis}) is reduced to a linear, first-order, 
nonhomogeneous equation, 
\begin{equation}
\label{nonhomo}
\frac{{\mathrm d}\phi}{{\mathrm d}x} - \left(\frac{\alpha}
{x + \lambda}\right)\phi = - \left(\frac{\eta \alpha}
{x + \lambda}\right) \,,
\end{equation}
in which $\eta$ plays the role of a nonhomogeneity parameter.

With $\mu =-1$ (implying a power-law in the distribution) and 
with $\alpha = -2$ (implying that the power-law is specifically
Zipf's law), the saturation properties of the network (for 
any value of $\eta$ and $\lambda$) can be abstracted 
from Eq.~(\ref{integ}) as
\begin{equation}
\label{finsol}
\phi (x) = \eta + \left(\frac{c}{x+\lambda}\right)^2 \,.
\end{equation}
One implication of the foregoing result 
is that nonhomogeneity in the system sets a firm lower bound to the 
number of rich nodes in the saturation regime, regardless of any 
arbitrarily high value of $x$, i.e. $\phi \longrightarrow \eta$
as $x \longrightarrow \infty$. In other words, nonhomogeneity 
defines a finite lower limit to the discrete count of the 
rich nodes. This clear deviation from the 
power-law model enables a few top nodes in the network of outgoing
links to get disproportionately rich, as shown in 
Figs.~\ref{f2} \&~\ref{f4}.
All the links from these top nodes are outwardly directed towards 
the dependent nodes, making the presence of these richly linked 
nodes an absolute necessity, and burdening them with the 
responsibility of maintaining functional coherence in the  
{\it FOSS} distribution. 

A scale for the onset of the saturation effects in the 
network of out-degree distributions can be found when  
the two terms in the right hand side of Eq.~(\ref{integ}) 
are in rough equipartition with each other. This will set a 
scale for the saturation of the number of links in the frequency 
distribution as
\begin{equation}
\label{satu}
x_{\mathrm{sat}} \simeq  
\frac{c}{\sqrt{{\vert \eta \vert}}} - \lambda \,.
\end{equation}
Considering the out-degree distribution in particular, it is 
always the case that $\eta =1$. From the fitting 
function, noting that $\lambda \ll c$, we can also conclude that 
$x_{\mathrm{sat}} \sim c$, a simple result that is useful in 
identifying the nodes which act as ``hubs" and deviate from
a scale-free distribution. All nodes with a number of links of
the order of $x_{\mathrm{sat}}$ or greater will belong to this
saturation regime. For the network of outgoing links, the 
{\it Debian} data indicate that approximately the top $1\%$ 
of the nodes
falls within this scale, with the package {\it libc6}  
being the most profusely connected node in all the releases.

The situation is quite the opposite for the network of incoming links,
as Figs.~\ref{f1} \&~\ref{f3} show. Here the nodes draw in links to 
themselves, with all links being inwardly directed towards the nodes. 
This network of incoming links is complementary in character to the 
network of outgoing links. As a result, the richly linked nodes of 
the latter network are poorly connected in the former. In contrast 
to Figs.~\ref{f2} \&~\ref{f4}, which indicate that 
the rich nodes serve the network to an extent that is disproportionately
greater than what a simple power-law behaviour would have required of
them, we see from Figs.~\ref{f1} \&~\ref{f3} that the most 
richly linked nodes in the in-degree distribution display a behaviour 
that falls short of what might be expected of a fully power-law trend 
(the top nodes here ought to have accreted more links if a power law 
only were to have been followed). So, decreasing values of $\eta$ 
over two generations of {\it Debian} releases, 
show that for a given number of links, $x$, the count of nodes, $\phi$,  
is reduced. It is then clear that 
the ability of the top nodes to acquire links in the in-degree
distribution becomes progressively
weakened (and so it is that the deviation from the power-law 
behaviour becomes sharper).
Saturation in the network can also be quantitatively determined
by the parameter $\eta$, which, when $\mu = -1$, appears as a
nonhomogeneity condition in Eq.~(\ref{logis}).
The value and especially the sign of $\eta$
afford us a precise means to distinguish
the directed network of incoming links from that of outgoing 
links. The difference in the respective degree distributions in
Figs.~\ref{f1} \&~\ref{f2} (or Figs.~\ref{f3} \&~\ref{f4}) 
underscores this fact.

For small values of $x$, the poorly linked nodes also deviate from
the power-law solution. This is especially true for the in-degree
distribution in Figs.~\ref{f1} \&~\ref{f3}. For small in-degree
and out-degree distributions in the World Wide Web~\cite{broku},
an improved fit is obtained by a simple modification in the
global power-law model~\cite{domen2,nsw}. This type of 
modification can also be engineered in Eq.~(\ref{finsol}) 
to obtain a similar fit for the weakly linked nodes.
In the limit of small degree distributions for both the in-directed
and out-directed networks, where $\eta$ ceases to have much
significance, and where $x \sim 1$ (which, in the discrete count of
links, is the lowest value that $x$ can assume practically), 
an upper bound to the number of the sparsely linked nodes is 
found to be 
\begin{equation}
\label{limphi}
\phi_{\mathrm{ub}} \simeq \left(\frac{c}{1 + \lambda}\right)^2 \,,
\end{equation}
with the full range of $\phi$, therefore, going as
$\eta \leq \phi \lesssim \phi_{\mathrm{ub}}$.

\section{Modelling evolution and saturation}
\label{sec4}

Our model, based on two generations ({\it Etch} and 
{\it Lenny}) of a standard {\it FOSS} network ({\it Debian}), has 
shown that the saturation properties of the in-degree and the 
out-degree distributions are differently affected as time passes 
(marked by new releases of {\it Debian}). 
The degree distribution of the network of outgoing links shows no 
change when it comes to the model fitting of the top nodes 
($\eta$ maintains the same value). This is  
expected of these nodes. They form the foundation of the whole 
network, and their prime status continues to hold.
In a semantic sense, meaning flows from these nodes to the 
derivative nodes. At the opposite end, 
the very poorly linked nodes in the outgoing 
network are fitted by changing values of $\lambda$ (as shown
in Figs.~\ref{f2} \&~\ref{f4}). Again this is expected. In a mature 
and robust network, the possibility of semantic variations is much 
more open in the weakly linked derivative nodes, as opposed to the 
primordial nodes.  

For the in-degree distributions, the situation is contrariwise.  
Going by Figs.~\ref{f1} \&~\ref{f3}, the model fitting can 
be achieved properly by changing the value of $\eta$ significantly.
Further, with a new release of {\it Debian}, $\eta$ actually decreases,
a fact whose import is that the most richly linked nodes in
the in-degree distribution (which are also the most dependent nodes)
acquire less links than what they might have done, if the power-law
trend were to have been adhered to indefinitely. So, from a dynamic
perspective, there is a limit upto which 
these dependent nodes continue to be linked. 

Taking these observations together, we realize  
that the {\it FOSS} network is not a static entity. Rather 
it is a dynamically evolving network, as any standard software network
is known to be~\cite{myers,gorpis}, undergoing continuous additions
(even deletions) and modifications across several generations of 
{\it Debian} releases, contributed by the community of free-software
developers. So any realistic model should account for this evolutionary
aspect of the network distribution, and by now many theoretical 
models~\cite{baral,krarele,domesa,krapred} have 
provided such insight into the 
dynamic evolution of networks. It is  
known too that scale-free networks emerge 
through the simultaneous operation of dynamic growth and preferential 
attachment~\cite{baral,baralje}. The limiting features of such a 
scale-free distribution ought also to come out naturally through 
the long-time dynamics. 

The top nodes in the out-degree
distribution form the irreducible nucleus of the {\it FOSS} network. 
These nodes are the most influential in the network. From the 
perspective of a continuum model, we look at the frequency 
distribution of the nodes in the network of outgoing links as a 
field, $\phi (x,t)$, evolving continuously through 
time, $t$, with the saturation in the number of nodes for high values
of $x$, emerging of its own accord from the dynamics. In keeping with
this need, we frame an ansatz with a general power-law feature as
\begin{equation}
\label{ansatz}
\phi (x,t) =\left(\frac{x + \lambda}{c}\right)^{\alpha} 
+ \varphi (x,t) \,,
\end{equation} 
in which $\varphi \longrightarrow \eta$, as $t \longrightarrow \infty$. 
This prescription is compatible with what Eq.~(\ref{integ}) 
indicates when $\mu = -1$. Under this requirement, 
the temporal evolution of the network is described by a  
first-order, linear, nonhomogeneous equation, going as,
\begin{equation}
\label{partpow}
\tau\frac{\partial \phi}{\partial t}= \frac{\partial \phi}{\partial x}
-\frac{\alpha}{c^{\alpha}}\left(x+\lambda\right)^{\alpha -1} \,,
\end{equation}
in which $\tau$ is a representative time 
scale on which the {\it FOSS} network evolves appreciably. 
Now, Eq.~(\ref{partpow}) already has 
a power-law property built in it explicitly, and is expected, 
upon being integrated under suitable initial conditions, to make
the saturation features of the top nodes appear because of 
nonhomogeneity. This is the exact reverse of Eq.~(\ref{nonhomo}), which 
has nonhomogeneity explicitly designed in it, and upon being integrated, 
leads to a power-law behaviour. 
The general solution of Eq.~(\ref{partpow}) can be 
obtained by the method of characteristics~\cite{ld97}, 
in which we need to solve the equations, 
\begin{equation}
\label{char}
- \frac{{\mathrm d}t}{\tau} = \frac{{\mathrm d}x}{1} = 
\frac{{\mathrm d}\phi}{\alpha(x+\lambda)^{\alpha -1}c^{-\alpha}} \,.
\end{equation}
The solution of the ${\mathrm d}\phi/{\mathrm d}x$ equation is 
\begin{equation}
\label{phiex}
\phi - \left(\frac{x + \lambda}{c}\right)^{\alpha} = a \,,
\end{equation}
while the solution of the 
${\mathrm d}x/{\mathrm d}t$ equation is
\begin{equation}
\label{ettee}
x + \frac{t}{\tau} = b \,,
\end{equation}
with both $a$ and $b$ being 
integration constants. The general solution is to be 
found under the condition that one characteristic solution 
of Eq.~(\ref{char}) is an arbitrary function of the other, i.e. 
$a = f(b)$, with $f$ having to be determined from the initial 
conditions~\cite{ld97}. So, going by the integral solutions 
given by Eqs.~(\ref{phiex}) and~(\ref{ettee}), the general solution 
of $\phi (x,t)$ will be  
\begin{equation}
\label{gensol}
f\left( x + \frac{t}{\tau} \right) = 
\phi - \left(\frac{x+\lambda}{c}\right)^{\alpha} \,,
\end{equation} 
which, under the initial condition that $\phi = \eta$ at $t=0$,  
will characterize the profile of the arbitrary function, $f$, as 
\begin{equation}
\label{arbfunc}
f(z) = \eta - \left(\frac{z + \lambda}{c}\right)^{\alpha} \,.
\end{equation}
Hence, the 
specific solution can be obtained from Eq.~(\ref{gensol}) as
\begin{equation}
\label{partsol1}
\phi (x,t)= \eta + \left(\frac{x+\lambda}{c}\right)^{\alpha} -
\left[\frac{1}{c}\left(x+\lambda+\frac{t}{\tau}\right)\right]^{\alpha} \,,
\end{equation}
and this, under the condition that $\alpha = -2$, will converge 
to the distribution given by Eq.~(\ref{finsol}), for 
$t \longrightarrow \infty$. The significance of the initial condition 
is worth stressing here. For a value of $x$, the evolution starts 
at $t=0$ with an initial node count of $\phi = \eta$, which, under 
all practical circumstances, is set at $\eta =1$. This is to 
say that a node appears in the network with $x$ number of links, 
where, previously, there existed no node with this particular 
number of links. As the network evolves, two things continue 
to happen: first, new nodes are added to the network, and secondly, 
already existing nodes accrete links in greater numbers. The most
heavily linked among the latter started as the primary
nodes, and at $t=0$, their number defines the minimum number
of independent packages that are absolutely necessary for the
{\it FOSS} network to evolve subsequently (for $t>0$) into a 
robust semantic system. From a semantic perspective, 
the initial condition can be argued
to have an axiomatic character, and the mature network burgeons
from it on later time scales. And during the evolution, the 
entire network gets dynamically self-organized in such a manner, 
that the eventual static out-degree distribution has its 
saturation properties at the highly connected end 
determined by what the initial field was like at $t=0$. 

The asymptotic properties of Eq.~(\ref{partsol1}) can now be 
examined, both in the limit of 
$t \longrightarrow 0$ and in the limit of $t \longrightarrow \infty$. 
In the former case, the evolution of $\phi$ will be linear in $t$
for a given value of $x$, and will go as  
\begin{equation}
\label{tee0}
\phi (x,t) \simeq \eta - \alpha 
\frac{\left(x + \lambda\right)^{\alpha -1}}{c^\alpha} 
\left(\frac{t}{\tau}\right) \,,
\end{equation}
in which growth is assured only when $\alpha < 0$. 
This linearity of early growth reflects the 
assumption of a linear growth of the number of nodes with 
time~\cite{valsol2}. 

While the temporal evolution obeys linearity on early time scales, 
in the opposite limit of $t \longrightarrow \infty$, the evolution 
shifts asymptotically to a power-law trend going as 
\begin{equation}
\label{teeinfty}
\phi (x,t) - \eta - \left(\frac{x + \lambda}{c}\right)^{\alpha}
\simeq - \frac{1}{c^\alpha}\left(\frac{t}{\tau}\right)^{\alpha} \,.
\end{equation}
Naturally, convergence towards a steady state, as it has been given 
by the condition in the left hand side of the foregoing relation, 
will be possible only when $\alpha < 0$, a requirement that is 
satisfied by Zipf's law ($\alpha = -2$). 
Free and open-source software has 
been known to have its dynamic processes driven by power 
laws~\cite{chaldu,chalval}, which is a clear sign that long memory 
prevails in this kind of a system. 

Now from the steady state form of the degree distribution, as 
it is given by Eq.~(\ref{integ}), we can set down, for
$\mu =-1$, a similar relation for the time-dependent field, 
$\phi \equiv \phi (x,t)$, as, 
\begin{equation}
\label{renor}
\phi (x,t) = \eta 
+ \left(\frac{x + {\tilde{\lambda}}}{\tilde{c}}\right)^{\alpha} \,,
\end{equation}
where $\tilde{\lambda}$ and $\tilde{c}$ are ``dressed" parameters,
defined as ${\tilde{\lambda}} = \lambda \nu (x,t)$ and     
${\tilde{c}} = c \zeta (x,t)$, respectively. The scaling form
of the two functions $\nu$ and $\zeta$ can be determined by equating
the right hand sides of Eqs.~(\ref{partsol1}) and~(\ref{renor}). 
This will lead to 
\begin{equation}
\label{scaleq}
\left(\frac{x + {\tilde{\lambda}}}{\tilde{c}}\right)^{\alpha} = 
\left(\frac{x+\lambda}{c}\right)^{\alpha} -
\left[\frac{1}{c}\left(x+\lambda+\frac{t}{\tau}\right)\right]^{\alpha} \,.
\end{equation}
For scales of $x \gg \lambda$ (typically $x \gtrsim 10$), a 
converging power-series
expansion of increasingly higher orders of $\lambda/x$ can be carried 
out with the help of Eq.~(\ref{scaleq}). The zeroth-order condition will 
deliver the scaling profile of $\zeta$ as 
\begin{equation}
\label{defzeta}
\zeta (x,t) = \left[1 - \left(1 + 
\frac{t}{x \tau}\right)^{\alpha}\right]^{-1/\alpha} \,.
\end{equation}   
This function bears time-translational properties, and at a given 
scale of $x$, it causes the degree distribution to shift across the 
$x$--$\phi$ plot through time. However, it is also not difficult 
to see that when $\alpha = -2$, there is a convergence towards $\zeta =1$
(the steady state limit) as $t \longrightarrow \infty$. 
And when $x \longrightarrow \infty$, on any finite time scale, 
$\zeta \longrightarrow 0$. This is why the count of the most
heavily connected nodes (for which $x$ has a high value) stays nearly
the same ($\phi = \eta$) at all times, a fact that is 
borne out by the out-degree distributions in Figs.~\ref{f2} \&~\ref{f4}.  
The saturation scale of $x$ for such behaviour is given 
by Eq.~(\ref{satu}). A related fact that also emerges is that 
time-translation of the degree distribution becomes steadily more 
pronounced
as we move away from $x \sim x_{\mathrm{sat}}$ towards the 
lower limit of $x =1$ (the least number of links that a node can 
possess). Consequently, as the temporal evolution progresses, the 
out-degree distribution assumes a slanted appearance with a negative
slope in the $x$--$\phi$ plane, something shown 
clearly in Figs.~\ref{f2} \&~\ref{f4}. 
The model fitting in these two
plots indicates that the value of $c$ increases with time. This is 
how it should be, going by the form of the scaling function
$\zeta (x,t)$, if we are careful to observe that $c$ in both 
the plots is to be viewed as $\tilde{c}$, to account for 
its time-dependent variation.   

Information regarding the time-translational properties of the 
poorly connected nodes is contained in the scaling 
function $\nu (x,t)$. However, a look at the left hand side of 
Eq.~(\ref{scaleq}) reveals that $\nu$ is coupled to $\zeta$, and
this nonlinear coupling causes complications. Going back 
to the power-series expansion in $\lambda/x$, as it can be obtained
from Eq.~(\ref{scaleq}), we may suppose that just as the
zeroth-order in the series has yielded a proper 
scaling form for $\zeta$, the higher orders in 
the series will bring forth a similar form for $\nu$. And indeed 
we do obtain such a solution, going as $\nu^k=\zeta^\alpha 
\left[1-\left(1+t/x\tau\right)^{\alpha -k}\right]$, with $k$ being
the order of the expansion in the power series. But this result is 
misleading because the parameter $\lambda$, and 
the scaling function $\nu (x,t)$, are influential only when
$\lambda \gtrsim x$, with $x$ assuming arbitrarily small values in the 
continuum model. Therefore, the correct approach here is not to take
a series expansion in $\lambda/x$, but rather in $x/\lambda$, with a 
proper convergence of the series taking place for higher orders in 
$x/\lambda$. The zeroth-order term of this series gives the scaling
form $\nu^\alpha = \zeta^\alpha 
\left[1 - \left(1 + t/\lambda \tau\right)^\alpha \right]$. The primary
difficulty with this result is that the true functional dependence of 
$\zeta$ in this case is not known. This is certainly not going to be 
the function that is implied by Eq.~(\ref{defzeta}), because this 
form of $\zeta$ is valid only on scales where $x \gg \lambda$.

\begin{figure}[floatfix]
\begin{center}
\includegraphics[scale=0.65, angle=0]{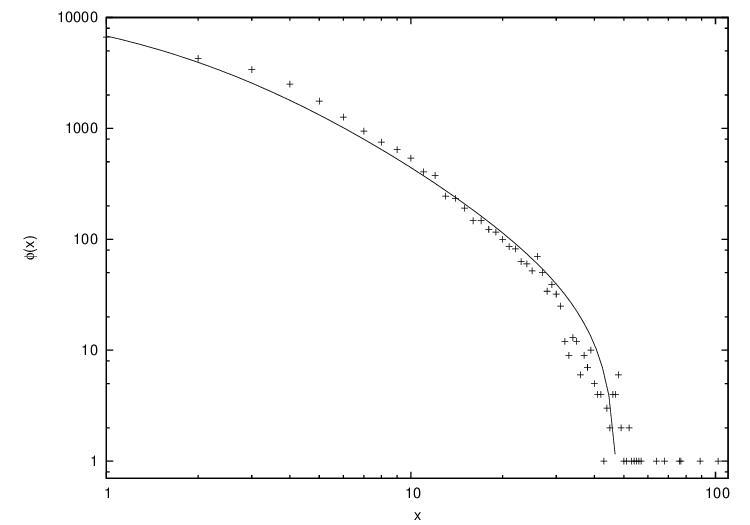}
\caption{\label{f5}\small{
On large scales of $x$ in the network of incoming links of the
latest stable release, {\it Squeeze}, the saturation of the
degree distribution is
fitted by the parameter value, $\eta = -28$.
When $x$ is small, the fit is obtained for $\lambda = 2.2$.
For this plot, $c \simeq 265$.}}
\end{center}
\end{figure}

\begin{figure}[floatfix]
\begin{center}
\includegraphics[scale=0.65, angle=0]{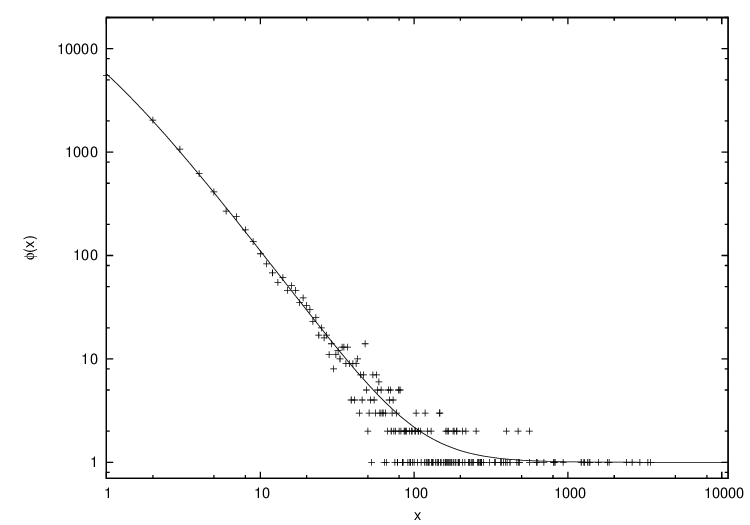}
\caption{\label{f6}\small{The out-degree distribution of the latest
stable release, {\it Squeeze}, is in agreement with what the dynamic
model predicts. The values of $\lambda$ and $c$ increase, as expected,
to $\lambda = 0.45$ and $c \simeq 110$. The richest
node in this distribution has $12470$ links.}}
\end{center}
\end{figure}

Considering everything, the clear message derived from the common 
pattern exhibited by the two generations of out-degree distributions 
is that the value of $\lambda$ has 
a significant bearing on the number of the preponderant but 
sparsely connected nodes, a fact that is described by 
Eq.~(\ref{limphi}). In the continuum picture of the degree distribution,
$\lambda$ is the theoretical lower bound of the number of links that 
the most weakly linked nodes may possess (which saves $\phi$ from 
suffering a divergence as $x \longrightarrow 0$, as 
Eq.~(\ref{finsol}) shows). Through the evolutionary growth of the
network, an increase in the value of $\lambda$ suggests that these
poorly linked nodes become incrementally relevant to the system by
contributing more links in the out-directed network. Now, these  
poorly connected nodes in the out-degree distribution are also 
the most profusely linked nodes in the in-directed network. 
Figs.~\ref{f1} \&~\ref{f3} show that for these nodes the value 
of $\eta$ decreases with the evolution of the {\it FOSS} network. 
So, while these nodes 
become progressively more relevant as members of the out-directed
network (a condition quantified by increasing values of $\lambda$),
as members of the in-directed network they become progressively 
less dependent (quantified by decreasing values of $\eta$). 
Analysing the data of all the six generations of 
{\it Debian}, it is seen for the out-directed network that 
the value of $\lambda$ remains nearly the same 
upto the fourth release, {\it Etch}, but grows noticeably thereafter 
for the next two releases, {\it Lenny} 
and {\it Squeeze}. 
Figs.\ref{f5} \&~\ref{f6} show, respectively, the in-degree and 
the out-degree distributions of the release,
{\it Squeeze}. 

In contrast, in the in-directed network, the value 
of $\lambda$ grows quickly for the early releases 
and then saturates 
in the {\it Lenny} and {\it Squeeze}
releases. Remembering that in the in-directed network the most poorly 
linked nodes are actually the parent nodes of the entire network, 
we conclude that even these
nodes become dependent on other nodes to a small extent. Taken as a 
whole, as time increases, the interdependency character of
the entire network becomes more firmly established, with even the 
relatively unimportant nodes showing a tendency to contribute 
outwardly-directed links.

Quantitative support in favour of this claim comes from the dynamics
of the out-directed network. In this case, the total number of nodes, 
$N_{\mathrm{out}}(t)$, at any given point of time, $t$, can be obtained 
by evaluating the integral
\begin{equation}
\label{summing}
N_{\mathrm{out}}(t) = \int^{x_{\mathrm{m}}}_1 \phi (x,t)\, {\mathrm d}x \,.
\end{equation}
The limits of this integral are decided by the limits on the number of 
links that the nodes possess, $1$ being the lower limit 
and $x_{\mathrm{m}}$ being the upper (maximum) limit. The integral in 
Eq.~(\ref{summing}) can be solved by taking the profile of $\phi (x,t)$ 
given by Eq.~(\ref{partsol1}), for $\alpha = -2$. Noting that  
$x_{\mathrm{m}}\gg 1$ (typically $x_{\mathrm{m}} \sim 10^4$) for the
out-directed network, the total number of nodes at any
time can be estimated as, 
\begin{equation}
\label{nodenum}
N_{\mathrm{out}}(t) \simeq \eta x_{\mathrm{m}}+\frac{c^2}{1+\lambda} 
- c^2 \left(1 + \lambda + \frac{t}{\tau}\right)^{-1} \,.
\end{equation}
On moderate time scales, the last two terms in the right hand side of 
Eq.~(\ref{nodenum}) are roughly equal. So the dominant contribution 
comes from the first term (the saturation term), as a consequence of 
which, we can set down $N_{\mathrm{out}} \sim x_{\mathrm{m}}$.
This argument becomes progressively more correct for large values 
of $x_{\mathrm{m}}$, i.e. for later releases of {\it Debian}.  

For the out-degree distribution in the {\it Etch} release, 
$x_{\mathrm{m}} \simeq 9000$, while in the {\it Lenny} release, 
the corresponding number is about $10000$. Using these values 
from both the releases of {\it Debian}, the respective count 
of $N_{\mathrm{out}}$ can be made for the two successive
generations. These values of $N_{\mathrm{out}}$
represent the number of nodes that contribute at least one 
link in the out-directed network.
In the case of the {\it Etch} release, the number of software 
packages contributing to the out-directed network is counted 
to be about $8700$ (which is closely comparable
to the estimated value of 
$N_{\mathrm{out}} \sim x_{\mathrm{m}} \simeq 9000$), 
and in the case of the {\it Lenny} release, the total count 
of the out-directed nodes is about $11000$ (which can be 
favourably compared once again to 
$N_{\mathrm{out}} \sim x_{\mathrm{m}} \simeq 10000$). 
As a fraction of the total number of nodes, these actual 
counts indicate that the number of nodes in the out-directed
network increases by $0.3\%$ from the {\it Etch} release to 
the {\it Lenny}
release. This validates the contention that 
with each passing generation, the network becomes incrementally  
more robust in terms of out-degree contributions coming from an 
increasingly greater number of nodes. 
The values pertaining to the latest stable release {\it Squeeze}, 
also go along with this trend. In this case the actual count 
of the out-directed nodes is about $14000$, a number that is 
again comparable with the estimate of 
$N_{\mathrm{out}} \sim x_{\mathrm{m}} \simeq 12000$. In keeping
with the predicted trend, 
the fraction of nodes contributing out-directed links in this
release increases by $1.2\%$. 
We also note with curiosity that in these last three 
{\it Debian} releases, {\it Etch}, {\it Lenny} and 
{\it Squeeze}, the total number of software packages, in both
the in-directed and out-directed networks, is roughly twice the 
value of $x_{\mathrm{m}}$ in the out-directed network. 
\begin{figure}[floatfix]
\begin{center}
\includegraphics[scale=0.65, angle=0]{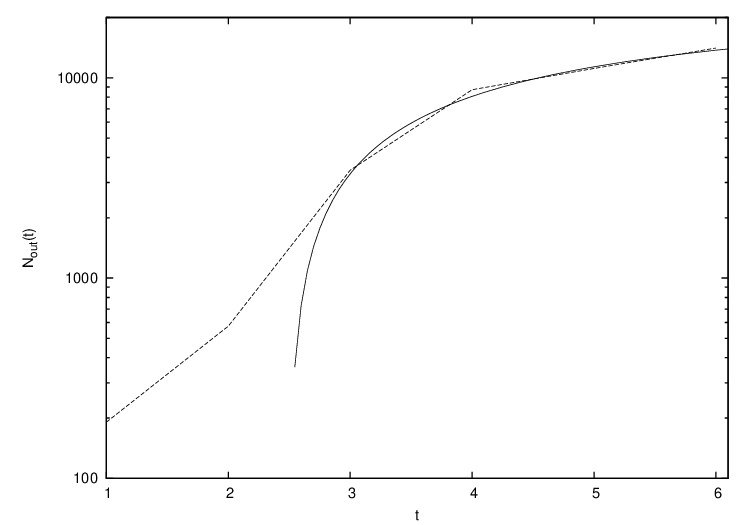}
\caption{\label{f7}\small{The broken dotted curve plots the growth
of the total number of nodes in the out-directed network over
six generations of {\it Debian}. The continuous curve, following
Eq.~(\ref{powasym}), gives the fitting function of the data
points. The fit, indicating a power-law approach towards
saturation, agrees well for the later releases of {\it Debian}
(the third release onwards). In this plot $t$ is a scaled time, 
marking the generation number. The parameter values are, 
$A=29000$, $B=113000$ and $C=1.4$, which are compatible
when viewed in terms of $\eta$, $\lambda$, $c$ and $x_{\mathrm{m}}$,
as they have been set in Figs.~\ref{f2},~\ref{f4} \&~\ref{f6}.}}
\end{center}
\end{figure}

The overall growth of the network, however, slowly grinds to a 
halt on long time scales. This conclusion cannot be missed in
Eq.~(\ref{nodenum}), which suggests that the total number of nodes 
increases with time, but approaches a finite stationary value when 
$t \longrightarrow \infty$, with $x_{\mathrm{m}}$ remaining finite. 
This inclination of the network to saturate towards a finite-sized 
end can be explained in terms of the finite
semantic possibilities associated with each of the nodes. The extent
of making creative use of the existing semantic possibilities of 
even the most intensely linked of the top nodes is limited. Since 
most of the nodes in the network depend on such top nodes, there 
must then be a terminal stage in the growth of the network. 
Unless novel creative elements in semantic terms are continuously 
added to the top nodes, the value of $x_{\mathrm{m}}$ will remain 
constrained within a certain bound, and saturation will happen. 
So saturation in the network is a consequence of the 
limit to the various ways in which original functions in 
the top nodes can be invoked by the derivative nodes. An 
illustration of this argument is to be seen in 
Fig.~\ref{f7}, which plots actual values taken from all the 
{\it Debian} releases. This plot tracks the growth
of the total number of nodes in the out-directed
network. All the members of this network contribute at least one 
out-directed link, and so meaning (the semantic context) is seen to
flow out of these nodes. Therefore, these nodes are the bearers of 
original axioms. That the growth of this entire out-directed network
saturates towards a limiting value for the later releases of 
{\it Debian} is quite 
obvious from the trend indicated in Fig.~\ref{f7}. The data curve
is fitted by a general form of Eq.~(\ref{nodenum}), going as
\begin{equation}
\label{powasym}
N_{\mathrm{out}}(t) \simeq A 
- \frac{B}{C +t} \,,
\end{equation}
where $A=\eta x_{\mathrm{m}}+c^2(1+\lambda)^{-2}$, $B=c^2\tau$
and $C=(1+\lambda)\tau$. The implication of the foregoing 
expression is that the long-time approach towards the 
terminal stage in the growth of the network is like a power law. 
From the fitting function, this looks very much true
for the later releases of {\it Debian}.
Now, saturation in the growth of the 
axiom-bearing nodes (with out-directed links) means 
that the growth of the network of in-directed nodes will also 
saturate in tandem. 
The semantic flow in the entire network terminates at these 
nodes, and as such these ``terminal" nodes 
are also indicators of saturation. 

\section{Concluding remarks}
\label{sec5}

This work is based on the networks of direct
dependencies in the component-based software, {\it Debian}. 
A deeper understanding of dependency-based semantic features
can be had on introducing the notions of strong
dependencies and package sensitivity, which are  
instrumental in distinguishing transitive 
dependencies, and conjunctive and disjunctive 
dependencies~\cite{zack}. We note that         
direct and strong dependencies generally tend to be
correlated~\cite{zack}.
These features may have a bearing on  
redundancy in the operating system and its 
robustness against failure.
We may also mention in passing that network structures 
in component-based software are determined by specific 
fields, with ``Depends", which is the basis of this study, 
being just one of such fields (``Conflicts", for
instance, being another). A particular field may give rise
to specific features in the network, characteristic of itself
only. 

The mathematical model developed in this work, 
makes a quantitative distinction between the incoming and 
outgoing distribution in the {\it Debian GNU/Linux} network. 
Similar features are known to exist in the degree distribution
of other scale-free networks, and with the mathematical 
framework applied here, it should become possible 
to study the saturation properties and the specific 
directional characteristics of scale-free networks in
general.  
To take an example, the degree distributions of the 
World Wide Web and {\it Debian} appear to be the 
converse of each other. And so what looks like an in-degree 
distribution for one, is the out-degree distribution
for the other, and vice versa~\cite{domen2}.  
The model provided here is general enough to capture the 
specific features of the two different cases by a suitable 
tuning of the parameters. 

\begin{acknowledgments}
We thank J. K. Bhattacharjee, A. Kumar, 
P. Majumdar, V. M. Yakovenko and S. Zacchiroli 
for helpful remarks. 
\end{acknowledgments}

\bibliography{nnr_2013}

\begin{thebibliography}{40}
\expandafter\ifx\csname natexlab\endcsname\relax\def\natexlab#1{#1}\fi
\expandafter\ifx\csname bibnamefont\endcsname\relax
  \def\bibnamefont#1{#1}\fi
\expandafter\ifx\csname bibfnamefont\endcsname\relax
  \def\bibfnamefont#1{#1}\fi
\expandafter\ifx\csname citenamefont\endcsname\relax
  \def\citenamefont#1{#1}\fi
\expandafter\ifx\csname url\endcsname\relax
  \def\url#1{\texttt{#1}}\fi
\expandafter\ifx\csname urlprefix\endcsname\relax\def\urlprefix{URL }\fi
\providecommand{\bibinfo}[2]{#2}
\providecommand{\eprint}[2][]{\url{#2}}

\bibitem[{\citenamefont{Dorogovtsev and Mendes}(2003)}]{domen2}
\bibinfo{author}{\bibfnamefont{S.~N.} \bibnamefont{Dorogovtsev}}
  \bibnamefont{and} \bibinfo{author}{\bibfnamefont{J.~F.~F.}
  \bibnamefont{Mendes}}, \emph{\bibinfo{title}{Evolution of Networks}}
  (\bibinfo{publisher}{Oxford University Press}, \bibinfo{address}{Oxford},
  \bibinfo{year}{2003}).

\bibitem[{\citenamefont{Newman et~al.}(2006)\citenamefont{Newman, Barab\'asi,
  and Watts}}]{nbw}
\bibinfo{author}{\bibfnamefont{M.}~\bibnamefont{Newman}},
  \bibinfo{author}{\bibfnamefont{A.-L.} \bibnamefont{Barab\'asi}},
  \bibnamefont{and} \bibinfo{author}{\bibfnamefont{D.~J.} \bibnamefont{Watts}},
  \emph{\bibinfo{title}{The Structure and Dynamics of Networks}}
  (\bibinfo{publisher}{Princeton University Press}, \bibinfo{address}{Princeton
  and Oxford}, \bibinfo{year}{2006}).

\bibitem[{\citenamefont{Barrat et~al.}(2008)\citenamefont{Barrat, Barth\'elemy,
  and Vespigniani}}]{bbv}
\bibinfo{author}{\bibfnamefont{A.}~\bibnamefont{Barrat}},
  \bibinfo{author}{\bibfnamefont{M.}~\bibnamefont{Barth\'elemy}},
  \bibnamefont{and}
  \bibinfo{author}{\bibfnamefont{A.}~\bibnamefont{Vespigniani}},
  \emph{\bibinfo{title}{Dynamical Processes on Complex Networks}}
  (\bibinfo{publisher}{Cambridge University Press},
  \bibinfo{address}{Cambridge}, \bibinfo{year}{2008}).

\bibitem[{\citenamefont{Dorogovtsev and Mendes}()}]{domen}
\bibinfo{author}{\bibfnamefont{S.~N.} \bibnamefont{Dorogovtsev}}
  \bibnamefont{and} \bibinfo{author}{\bibfnamefont{J.~F.~F.}
  \bibnamefont{Mendes}}, \eprint{cond-mat/0106144}.

\bibitem[{\citenamefont{Strogatz}(2001)}]{stronat}
\bibinfo{author}{\bibfnamefont{S.~H.} \bibnamefont{Strogatz}},
  \bibinfo{journal}{Nature} \textbf{\bibinfo{volume}{410}},
  \bibinfo{pages}{268} (\bibinfo{year}{2001}).

\bibitem[{\citenamefont{Albert and Barab\'asi}(2002)}]{albar}
\bibinfo{author}{\bibfnamefont{R.}~\bibnamefont{Albert}} \bibnamefont{and}
  \bibinfo{author}{\bibfnamefont{A.-L.} \bibnamefont{Barab\'asi}},
  \bibinfo{journal}{Rev. Mod. Phys.} \textbf{\bibinfo{volume}{74}},
  \bibinfo{pages}{47} (\bibinfo{year}{2002}).

\bibitem[{\citenamefont{Newman}(2003)}]{new}
\bibinfo{author}{\bibfnamefont{M.~E.~J.} \bibnamefont{Newman}},
  \bibinfo{journal}{SIAM Review} \textbf{\bibinfo{volume}{45}},
  \bibinfo{pages}{167} (\bibinfo{year}{2003}).

\bibitem[{\citenamefont{Amaral and Ottino}(2004)}]{amot}
\bibinfo{author}{\bibfnamefont{L.~A.~N.} \bibnamefont{Amaral}}
  \bibnamefont{and} \bibinfo{author}{\bibfnamefont{J.~M.}
  \bibnamefont{Ottino}}, \bibinfo{journal}{Eur. Phys. J. B}
  \textbf{\bibinfo{volume}{38}}, \bibinfo{pages}{147} (\bibinfo{year}{2004}).

\bibitem[{\citenamefont{Newman}()}]{markn}
\bibinfo{author}{\bibfnamefont{M.~E.~J.} \bibnamefont{Newman}},
  \eprint{arXiv:1112.1440}.

\bibitem[{\citenamefont{Albert et~al.}(1999)\citenamefont{Albert, Jeong, and
  Barab\'asi}}]{ajb}
\bibinfo{author}{\bibfnamefont{R.}~\bibnamefont{Albert}},
  \bibinfo{author}{\bibfnamefont{H.}~\bibnamefont{Jeong}}, \bibnamefont{and}
  \bibinfo{author}{\bibfnamefont{A.-L.} \bibnamefont{Barab\'asi}},
  \bibinfo{journal}{Nature} \textbf{\bibinfo{volume}{401}},
  \bibinfo{pages}{130} (\bibinfo{year}{1999}).

\bibitem[{\citenamefont{Chatterjee and Chakrabarti(Eds.)}(2007)}]{acbkc1}
\bibinfo{author}{\bibfnamefont{A.}~\bibnamefont{Chatterjee}} \bibnamefont{and}
  \bibinfo{author}{\bibfnamefont{B.~K.} \bibnamefont{Chakrabarti(Eds.)}},
  \emph{\bibinfo{title}{Econophysics of Markets and Business Networks}}
  (\bibinfo{publisher}{Springer-Verlag Italia}, \bibinfo{address}{Milano},
  \bibinfo{year}{2007}).

\bibitem[{\citenamefont{i~Cancho et~al.}(2004)\citenamefont{i~Cancho, Sol\'e,
  and K{\"o}hler}}]{cansolko}
\bibinfo{author}{\bibfnamefont{R.~F.} \bibnamefont{i~Cancho}},
  \bibinfo{author}{\bibfnamefont{R.~V.} \bibnamefont{Sol\'e}},
  \bibnamefont{and}
  \bibinfo{author}{\bibfnamefont{R.}~\bibnamefont{K{\"o}hler}},
  \bibinfo{journal}{Phys. Rev. E} \textbf{\bibinfo{volume}{69}},
  \bibinfo{pages}{051915} (\bibinfo{year}{2004}).

\bibitem[{\citenamefont{i~Cancho}(2004)}]{cancho}
\bibinfo{author}{\bibfnamefont{R.~F.} \bibnamefont{i~Cancho}},
  \bibinfo{journal}{Phys. Rev. E} \textbf{\bibinfo{volume}{70}},
  \bibinfo{pages}{056135} (\bibinfo{year}{2004}).

\bibitem[{\citenamefont{Steyvers and Tenenbaum}(2005)}]{stete}
\bibinfo{author}{\bibfnamefont{M.}~\bibnamefont{Steyvers}} \bibnamefont{and}
  \bibinfo{author}{\bibfnamefont{J.~B.} \bibnamefont{Tenenbaum}},
  \bibinfo{journal}{Cognitive Science: A Multidisciplinary Journal}
  \textbf{\bibinfo{volume}{29(1)}}, \bibinfo{pages}{41} (\bibinfo{year}{2005}).

\bibitem[{\citenamefont{i~Cancho et~al.}(2001)\citenamefont{i~Cancho, Janssen,
  and Sol\'e}}]{canjaso}
\bibinfo{author}{\bibfnamefont{R.~F.} \bibnamefont{i~Cancho}},
  \bibinfo{author}{\bibfnamefont{C.}~\bibnamefont{Janssen}}, \bibnamefont{and}
  \bibinfo{author}{\bibfnamefont{R.~V.} \bibnamefont{Sol\'e}},
  \bibinfo{journal}{Phys. Rev. E} \textbf{\bibinfo{volume}{64}},
  \bibinfo{pages}{046119} (\bibinfo{year}{2001}).

\bibitem[{\citenamefont{Valverde et~al.}(2002)\citenamefont{Valverde, Cancho,
  and Sol\'e}}]{valcaso}
\bibinfo{author}{\bibfnamefont{S.}~\bibnamefont{Valverde}},
  \bibinfo{author}{\bibfnamefont{R.~F.} \bibnamefont{Cancho}},
  \bibnamefont{and} \bibinfo{author}{\bibfnamefont{R.~V.}
  \bibnamefont{Sol\'e}}, \bibinfo{journal}{Europhys. Lett.}
  \textbf{\bibinfo{volume}{60}}, \bibinfo{pages}{512} (\bibinfo{year}{2002}).

\bibitem[{\citenamefont{Valverde and Sol\'e}()}]{valsol}
\bibinfo{author}{\bibfnamefont{S.}~\bibnamefont{Valverde}} \bibnamefont{and}
  \bibinfo{author}{\bibfnamefont{R.~V.} \bibnamefont{Sol\'e}},
  \eprint{cond-mat/0307278}.

\bibitem[{\citenamefont{Challet and Lombardoni}(2004)}]{challom}
\bibinfo{author}{\bibfnamefont{D.}~\bibnamefont{Challet}} \bibnamefont{and}
  \bibinfo{author}{\bibfnamefont{A.}~\bibnamefont{Lombardoni}},
  \bibinfo{journal}{Phys. Rev. E} \textbf{\bibinfo{volume}{70}},
  \bibinfo{pages}{046109} (\bibinfo{year}{2004}).

\bibitem[{\citenamefont{LaBelle and Wallingford}()}]{labwal}
\bibinfo{author}{\bibfnamefont{N.}~\bibnamefont{LaBelle}} \bibnamefont{and}
  \bibinfo{author}{\bibfnamefont{E.}~\bibnamefont{Wallingford}},
  \eprint{cs/0411096}.

\bibitem[{\citenamefont{Challet and Du}(2005)}]{chaldu}
\bibinfo{author}{\bibfnamefont{D.}~\bibnamefont{Challet}} \bibnamefont{and}
  \bibinfo{author}{\bibfnamefont{Y.~L.} \bibnamefont{Du}},
  \bibinfo{journal}{International Journal of Reliability, Quality and Safety
  Engineering} \textbf{\bibinfo{volume}{12}}, \bibinfo{pages}{521}
  (\bibinfo{year}{2005}).

\bibitem[{\citenamefont{Challet and Valverde}()}]{chalval}
\bibinfo{author}{\bibfnamefont{D.}~\bibnamefont{Challet}} \bibnamefont{and}
  \bibinfo{author}{\bibfnamefont{S.}~\bibnamefont{Valverde}},
  \eprint{arXiv:0802.3170}.

\bibitem[{\citenamefont{Watts and Strogatz}(1998)}]{watstro}
\bibinfo{author}{\bibfnamefont{D.~J.} \bibnamefont{Watts}} \bibnamefont{and}
  \bibinfo{author}{\bibfnamefont{S.~H.} \bibnamefont{Strogatz}},
  \bibinfo{journal}{Nature} \textbf{\bibinfo{volume}{393}},
  \bibinfo{pages}{440} (\bibinfo{year}{1998}).

\bibitem[{\citenamefont{Abate et~al.}(2009)\citenamefont{Abate, Cosmo, Boender,
  and Zacchiroli}}]{zack}
\bibinfo{author}{\bibfnamefont{P.}~\bibnamefont{Abate}},
  \bibinfo{author}{\bibfnamefont{R.~D.} \bibnamefont{Cosmo}},
  \bibinfo{author}{\bibfnamefont{J.}~\bibnamefont{Boender}}, \bibnamefont{and}
  \bibinfo{author}{\bibfnamefont{S.}~\bibnamefont{Zacchiroli}},
  \bibinfo{journal}{Empirical Software Engineering and Measurement}
  p.~\bibinfo{pages}{89} (\bibinfo{year}{2009}).

\bibitem[{\citenamefont{Maillart et~al.}(2008)\citenamefont{Maillart, Sornette,
  Spaeth, and von Krogh}}]{mssk}
\bibinfo{author}{\bibfnamefont{T.}~\bibnamefont{Maillart}},
  \bibinfo{author}{\bibfnamefont{D.}~\bibnamefont{Sornette}},
  \bibinfo{author}{\bibfnamefont{S.}~\bibnamefont{Spaeth}}, \bibnamefont{and}
  \bibinfo{author}{\bibfnamefont{G.}~\bibnamefont{von Krogh}},
  \bibinfo{journal}{Phys. Rev. Lett.} \textbf{\bibinfo{volume}{101}},
  \bibinfo{pages}{218701} (\bibinfo{year}{2008}).

\bibitem[{\citenamefont{Zipf}(1949)}]{zipf}
\bibinfo{author}{\bibfnamefont{G.~K.} \bibnamefont{Zipf}},
  \emph{\bibinfo{title}{Human Behavior and the Principle of Least Effort}}
  (\bibinfo{publisher}{Addison-Wesley Press Inc.}, \bibinfo{address}{Cambridge,
  Massachusetts}, \bibinfo{year}{1949}).

\bibitem[{\citenamefont{Barab\'asi and Stanley}(1995)}]{barstan}
\bibinfo{author}{\bibfnamefont{A.-L.} \bibnamefont{Barab\'asi}}
  \bibnamefont{and} \bibinfo{author}{\bibfnamefont{H.~E.}
  \bibnamefont{Stanley}}, \emph{\bibinfo{title}{Fractal Concepts in Surface
  Growth}} (\bibinfo{publisher}{Cambridge University Press},
  \bibinfo{address}{Cambridge}, \bibinfo{year}{1995}).

\bibitem[{\citenamefont{Newman et~al.}(2001)\citenamefont{Newman, Strogatz, and
  Watts}}]{nsw}
\bibinfo{author}{\bibfnamefont{M.~E.~J.} \bibnamefont{Newman}},
  \bibinfo{author}{\bibfnamefont{S.~H.} \bibnamefont{Strogatz}},
  \bibnamefont{and} \bibinfo{author}{\bibfnamefont{D.~J.} \bibnamefont{Watts}},
  \bibinfo{journal}{Phys. Rev. E} \textbf{\bibinfo{volume}{64}},
  \bibinfo{pages}{026118} (\bibinfo{year}{2001}).

\bibitem[{\citenamefont{Nair et~al.}(2012)\citenamefont{Nair, Nagarjuna, and
  Ray}}]{nnrjphys}
\bibinfo{author}{\bibfnamefont{R.}~\bibnamefont{Nair}},
  \bibinfo{author}{\bibfnamefont{G.}~\bibnamefont{Nagarjuna}},
  \bibnamefont{and} \bibinfo{author}{\bibfnamefont{A.~K.} \bibnamefont{Ray}},
  \bibinfo{journal}{J. Phys.: Conf. Ser.} \textbf{\bibinfo{volume}{365}},
  \bibinfo{pages}{012058} (\bibinfo{year}{2012}).

\bibitem[{\citenamefont{Montroll}(1978)}]{montroll}
\bibinfo{author}{\bibfnamefont{E.~W.} \bibnamefont{Montroll}},
  \bibinfo{journal}{Proceedings of the National Academy of Science of the USA}
  \textbf{\bibinfo{volume}{75}}, \bibinfo{pages}{4633} (\bibinfo{year}{1978}).

\bibitem[{\citenamefont{Strogatz}(1994)}]{stro}
\bibinfo{author}{\bibfnamefont{S.~H.} \bibnamefont{Strogatz}},
  \emph{\bibinfo{title}{Nonlinear Dynamics and Chaos}}
  (\bibinfo{publisher}{Addison-Wesley Publishing Company},
  \bibinfo{address}{Reading, MA}, \bibinfo{year}{1994}).

\bibitem[{\citenamefont{Broder et~al.}(2000)\citenamefont{Broder, Kumar,
  Maghoul, Raghavan, Rajagopalan, Stata, Tomkins, and Wiener}}]{broku}
\bibinfo{author}{\bibfnamefont{A.}~\bibnamefont{Broder}},
  \bibinfo{author}{\bibfnamefont{R.}~\bibnamefont{Kumar}},
  \bibinfo{author}{\bibfnamefont{F.}~\bibnamefont{Maghoul}},
  \bibinfo{author}{\bibfnamefont{P.}~\bibnamefont{Raghavan}},
  \bibinfo{author}{\bibfnamefont{S.}~\bibnamefont{Rajagopalan}},
  \bibinfo{author}{\bibfnamefont{R.}~\bibnamefont{Stata}},
  \bibinfo{author}{\bibfnamefont{A.}~\bibnamefont{Tomkins}}, \bibnamefont{and}
  \bibinfo{author}{\bibfnamefont{J.}~\bibnamefont{Wiener}},
  \bibinfo{journal}{Comput. Netw.} \textbf{\bibinfo{volume}{33}},
  \bibinfo{pages}{309} (\bibinfo{year}{2000}).

\bibitem[{\citenamefont{Myers}(2003)}]{myers}
\bibinfo{author}{\bibfnamefont{C.~R.} \bibnamefont{Myers}},
  \bibinfo{journal}{Phys. Rev. E} \textbf{\bibinfo{volume}{68}},
  \bibinfo{pages}{046116} (\bibinfo{year}{2003}).

\bibitem[{\citenamefont{Gorshenev and Pis'mak}(2004)}]{gorpis}
\bibinfo{author}{\bibfnamefont{A.~A.} \bibnamefont{Gorshenev}}
  \bibnamefont{and} \bibinfo{author}{\bibfnamefont{Y.~M.}
  \bibnamefont{Pis'mak}}, \bibinfo{journal}{Phys. Rev. E}
  \textbf{\bibinfo{volume}{70}}, \bibinfo{pages}{067103}
  (\bibinfo{year}{2004}).

\bibitem[{\citenamefont{Barab\'asi and Albert}(1999)}]{baral}
\bibinfo{author}{\bibfnamefont{A.-L.} \bibnamefont{Barab\'asi}}
  \bibnamefont{and} \bibinfo{author}{\bibfnamefont{R.}~\bibnamefont{Albert}},
  \bibinfo{journal}{Science} \textbf{\bibinfo{volume}{286}},
  \bibinfo{pages}{509} (\bibinfo{year}{1999}).

\bibitem[{\citenamefont{Krapivsky et~al.}(2000)\citenamefont{Krapivsky, Redner,
  and Levyraz}}]{krarele}
\bibinfo{author}{\bibfnamefont{P.~L.} \bibnamefont{Krapivsky}},
  \bibinfo{author}{\bibfnamefont{S.}~\bibnamefont{Redner}}, \bibnamefont{and}
  \bibinfo{author}{\bibfnamefont{F.}~\bibnamefont{Levyraz}},
  \bibinfo{journal}{Phys. Rev. Lett.} \textbf{\bibinfo{volume}{85}},
  \bibinfo{pages}{4629} (\bibinfo{year}{2000}).

\bibitem[{\citenamefont{Dorogovtsev et~al.}(2000)\citenamefont{Dorogovtsev,
  Mendes, and Samukhin}}]{domesa}
\bibinfo{author}{\bibfnamefont{S.~N.} \bibnamefont{Dorogovtsev}},
  \bibinfo{author}{\bibfnamefont{J.~F.~F.} \bibnamefont{Mendes}},
  \bibnamefont{and} \bibinfo{author}{\bibfnamefont{A.~N.}
  \bibnamefont{Samukhin}}, \bibinfo{journal}{Phys. Rev. Lett.}
  \textbf{\bibinfo{volume}{85}}, \bibinfo{pages}{4633} (\bibinfo{year}{2000}).

\bibitem[{\citenamefont{Krapivsky and Redner}(2005)}]{krapred}
\bibinfo{author}{\bibfnamefont{P.~L.} \bibnamefont{Krapivsky}}
  \bibnamefont{and} \bibinfo{author}{\bibfnamefont{S.}~\bibnamefont{Redner}},
  \bibinfo{journal}{Phys. Rev. E} \textbf{\bibinfo{volume}{71}},
  \bibinfo{pages}{036118} (\bibinfo{year}{2005}).

\bibitem[{\citenamefont{Barab\'asi et~al.}(1999)\citenamefont{Barab\'asi,
  Albert, and Jeong}}]{baralje}
\bibinfo{author}{\bibfnamefont{A.-L.} \bibnamefont{Barab\'asi}},
  \bibinfo{author}{\bibfnamefont{R.}~\bibnamefont{Albert}}, \bibnamefont{and}
  \bibinfo{author}{\bibfnamefont{H.}~\bibnamefont{Jeong}},
  \bibinfo{journal}{Physica A} \textbf{\bibinfo{volume}{272}},
  \bibinfo{pages}{173} (\bibinfo{year}{1999}).

\bibitem[{\citenamefont{Debnath}(1997)}]{ld97}
\bibinfo{author}{\bibfnamefont{L.}~\bibnamefont{Debnath}},
  \emph{\bibinfo{title}{Nonlinear Partial Differential Equations for Scientists
  and Engineers}} (\bibinfo{publisher}{Birkh{\"a}user},
  \bibinfo{address}{Boston}, \bibinfo{year}{1997}).

\bibitem[{\citenamefont{Valverde and Sol\'e}(2005)}]{valsol2}
\bibinfo{author}{\bibfnamefont{S.}~\bibnamefont{Valverde}} \bibnamefont{and}
  \bibinfo{author}{\bibfnamefont{R.~V.} \bibnamefont{Sol\'e}},
  \bibinfo{journal}{Europhys. Lett.} \textbf{\bibinfo{volume}{72(5)}},
  \bibinfo{pages}{858} (\bibinfo{year}{2005}).

\end{thebibliography}

\end{document}